\documentclass[aps,pra,superscriptaddress]{revtex4-2}

\usepackage{amsmath,amsfonts,amssymb}
\usepackage{xcolor,graphicx}
\usepackage{braket}
\usepackage{float}
\usepackage[pdftex]{hyperref}
\hypersetup{
	colorlinks = true,
	linkcolor = blue,
	anchorcolor = blue,
	citecolor = red,
	filecolor = blue,
	urlcolor = blue}
\usepackage{subfig}
\usepackage{overpic}

\usepackage{listings}
\usepackage{xcolor}
\usepackage[utf8]{inputenc}
\usepackage{float}

\lstdefinestyle{python}{
    language=Python,
    backgroundcolor=\color{white},   
    basicstyle=\footnotesize\ttfamily, 
    breaklines=true,                  
    captionpos=b,                     
    numbers=left,                     
    numberstyle=\tiny\color{gray},   
    keywordstyle=\color{blue},        
    commentstyle=\color{green!80!black}, 
    stringstyle=\color{red},          
}

\begin{document}

\title{Numerical study of the radiation-matter interaction quantum systems through the time-dependent Schrödinger dynamics}

\author{L. Hernández-Sánchez}
\email[e-mail: ]{leonardi1469@gmail.com}
\affiliation{Instituto Nacional de Astrofísica Óptica y Electrónica, Calle Luis Enrique Erro No. 1\\ Santa María Tonantzintla, Puebla, 72840, Mexico}
\author{I. Ramos-Prieto}
\affiliation{Instituto Nacional de Astrofísica Óptica y Electrónica, Calle Luis Enrique Erro No. 1\\ Santa María Tonantzintla, Puebla, 72840, Mexico}
\author{A. Flores-Rosas}
\affiliation{Facultad de Ciencias en F\'isica y Matem\'aticas, Universidad Aut\'onoma de Chiapas, Carretera Emiliano Zapata, Km. 8, Rancho San Francisco, 29050 Tuxtla Guti\'errez, Chiapas, Mexico}
\author{F. Soto-Eguibar}
\affiliation{Instituto Nacional de Astrofísica Óptica y Electrónica, Calle Luis Enrique Erro No. 1\\ Santa María Tonantzintla, Puebla, 72840, Mexico}
\author{H.M. Moya-Cessa}
\affiliation{Instituto Nacional de Astrofísica Óptica y Electrónica, Calle Luis Enrique Erro No. 1\\ Santa María Tonantzintla, Puebla, 72840, Mexico}

\date{\today}

\begin{abstract}
La obtención de soluciones exactas de la ecuación de Schrödinger en sistemas cuánticos complejos plantea importantes desafíos. En este contexto, los métodos numéricos se presentan como una herramienta valiosa para el análisis de dichos sistemas. Este artículo propone un enfoque numérico utilizando el método de Runge-Kutta de cuarto orden, implementado en Python, para abordar sistemas de interacción radiación-materia. Esta metodología es aplicable a diversos hamiltonianos, como el modelo de Jaynes-Cummings. La precisión de los resultados numéricos se valida mediante la comparación con soluciones analíticas en casos simplificados, lo que demuestra su eficacia en el estudio de sistemas cuánticos donde las soluciones exactas no pueden derivarse.\\
\textbf{Palabras clave}: Ecuación de Schrödinger, interacción radiación-materia, método de Runge-Kutta, modelo de Jaynes-Cummings, Python.
\begin{center}
    \textbf{Abstract}
\end{center}
Obtaining exact solutions to the Schrödinger equation in complex quantum systems poses significant challenges. In this context, numerical methods emerge as valuable tools for analyzing such systems. This article proposes a numerical approach using the fourth-order Runge-Kutta method, implemented in Python, to tackle radiation-matter interaction systems. This methodology is applicable to various Hamiltonians, including that of the Jaynes-Cummings model. The accuracy of the numerical results is validated by comparing them with analytical solutions in simplified cases, demonstrating its effectiveness in studying quantum systems where exact solutions cannot be derived.
\end{abstract}
\maketitle

\section{Introduction}
The study of interactions between radiation and matter is a fundamental aspect of quantum physics, with crucial applications in quantum optics and spectroscopy, among other fields~\cite{Gerry_Book, Klimov_2009, Garrison_2008, Agarwal_2013, Fox_2006,Lara}. A prominent example of these interactions is the Jaynes-Cummings model, which describes the interaction between a two-level atom and a quantized electromagnetic field in a single-mode cavity. This model is particularly valuable because it provides exact analytical solutions under certain approximations, such as the dipole approximation and the rotating wave approximation, thus facilitating the analysis of specific interactions under well-defined conditions~\cite{JC_1963, Bruce_1993, Larson_2022,Salado}.

In quantum optics, the study of radiation-matter interaction has led to the development of more complex generalizations of the Jaynes-Cummings model, encompassing multiple atomic levels~\cite{Buzek_1990, Nath_2003, Hernandez_2017}, electromagnetic field modes~\cite{Sukumar_1981, Abdalla_1991}, and interaction with external fields~\cite{Alsing_1992, Hernandez_book, Bocanegra_2023}, to mention only some; however, the complexity of these models makes obtaining exact solutions to the Schrödinger equation challenging. Although certain approximations or transformations can simplify Hamiltonians modeling these systems, they do not always produce a complete description~\cite{Gerry_Book, Amaro_2015}. Therefore, numerical methods are essential for analyzing these advanced quantum phenomena~\cite{Butcher_book, Press_book, Qutip}.

The fourth-order Runge-Kutta (RK4) method is a prominent numerical technique for solving all types of differential equations, including the Schrödinger equation. Its accuracy and versatility make it ideal for simulating the time evolution of quantum systems with complex interactions~\cite{Butcher_book, Press_book}. When implemented in programming languages such as Python, RK4 provides a robust and efficient approximation, aiding in the analysis and understanding of complex quantum systems~\cite{Qutip}.

This article presents a methodology based on RK4, implemented in Python, to solve the Schrödinger equation within the context of radiation-matter interactions. The goal is to deliver a numerical solution that exceeds the limitations of traditional analytical approaches, thus enabling the study of complex quantum phenomena. The structure of the paper is as follows: Section~\ref{Model} provides a detailed analysis of the most general radiation-matter interaction model, known as the Jaynes-Cummings model. Then, in Section~\ref{Methodology}, we develop step-by-step the methodology used to solve the Schrödinger equation associated with this Hamiltonian. These solutions are tested in Section~\ref{Resultados}, where we compare and analyze both the analytical and numerical solutions obtained through this methodology. Finally, in Section~\ref{Conclusions}, we present our conclusions.

\section{Description of a Quantum System of Radiation and Matter in Interaction}\label{Model}
\begin{figure}[H] 
    \centering 
    \includegraphics[width=0.5\linewidth]{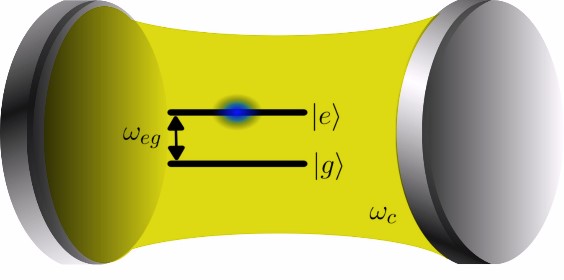}
    \caption{Representation of the Jaynes-Cummings model: a system composed of a two-level atom, where the excited state is denoted as $\ket{e}$ and the ground state as $\ket{g}$. This atom is confined within an optical cavity, which is made of perfectly reflective mirrors.}
    \label{Fig1}
\end{figure}

The Jaynes-Cummings model is fundamental for the analysis of quantum systems of radiation and matter in interaction~\cite{JC_1963}. This model describes the interaction between a two-level atom and a quantized electromagnetic field in a single-mode cavity, as illustrated in figure \ref{Fig1}. Its main advantage is that it allows for exact analytical solutions under certain approximations, making it a solid basis for studying more complex interactions~\cite{Bruce_1993,Larson_2022}.
Based on the dipole approximation and the rotating wave approximation, the Hamiltonian describing this system is expressed as~\cite{JC_1963, Gerry_Book}
\begin{equation} \label{Hamiltonian JCM}
    \hat{H} = \hbar \omega_c \hat{a}^\dagger \hat{a} + \frac{1}{2} \hbar \omega_{eg} \hat{\sigma}_z + \hbar g (\hat{\sigma}_+ \hat{a} + \hat{\sigma}_- \hat{a}^\dagger),
\end{equation}
where $\hbar$ is the reduced Planck's constant. The terms that comprise this Hamiltonian are physically interpreted as follows:
\begin{itemize}
    \item \textbf{Energy of the electromagnetic field}: The first term, $\hbar \omega_c \hat{a}^\dagger \hat{a}$, represents the energy of the electromagnetic field within the cavity. In this expression, $\hat{a}$ and $\hat{a}^\dagger$ are the annihilation and creation operators, respectively, and $\omega_c$ is the field's frequency.
    \item \textbf{Energy of the two-level atom}: The second term, $\frac{1}{2} \hbar \omega_{eg} \hat{\sigma}_z$, corresponds to the energy of the atom, where $\hat{\sigma}_z$ is the Pauli operator and $\omega_{eg}$ is the transition frequency between the two energy levels.
    \item \textbf{Atom-field interaction}: The third term, $\hbar g (\hat{\sigma}_+ \hat{a} + \hat{\sigma}_- \hat{a}^\dagger)$, describes the interaction between the electromagnetic field and the atom. In this expression, $g$ is the coupling constant that quantifies the strength of the interaction. This term has two components:
        \begin{itemize}
            \item \textbf{Photon absorption}: The part $\hat{\sigma}_+ \hat{a}$ describes the process in which the atom absorbs a photon and transitions from its ground state $\ket{g}$ to the excited state $\ket{e}$.
            \item \textbf{Photon emission}: The part $\hat{\sigma}_- \hat{a}^\dagger$ corresponds to the emission process, where the atom transitions from the excited state $\ket{e}$ to the ground state $\ket{g}$, emitting a photon in the process.
        \end{itemize}
\end{itemize}

The system's temporal evolution is described by the Schrödinger equation
\begin{equation}\label{Ecuacion de Schrodinger}
    i \hbar \frac{\partial}{\partial t} |\psi(t)\rangle = \hat{H} |\psi(t)\rangle,
\end{equation}
where $|\psi(t)\rangle$ represents the quantum state of the system at time $t$. The solution to this equation provides a complete description of the system's evolution.

For Hamiltonians who model more complex interactions, finding exact solutions can be extremely challenging and often impossible. Although approximations or transformations can be used, they do not always provide a complete analysis. In such circumstances, numerical methods become essential to overcome these limitations and obtain a more precise understanding of the system.

\section{Methodology}\label{Methodology}

In this chapter, RK4 is presented as an effective tool to solve the Schrödinger equation~\eqref{Ecuacion de Schrodinger} associated with the Hamiltonian of the Jaynes-Cummings model~\eqref{Hamiltonian JCM}. This approach is applicable not only to this particular model but can also be extended to more complex Hamiltonians, where analytical solutions are unattainable.

\subsection{Problem statement}
As previously mentioned, the time evolution of the quantum state $|\psi(t)\rangle$ is governed by the Schrödinger equation~\eqref{Ecuacion de Schrodinger} associated with the system Hamiltonian~\eqref{Hamiltonian JCM}. The main goal of this work is to develop a methodology based on RK4, implemented in Python, to solve this equation and obtain the wave function $|\psi(t)\rangle$ as a function of time. Subsequently, this approach will be generalized to be applied to other Hamiltonians modeling more complex systems.

Furthermore, this methodology will allow simulating the system's dynamics under various initial conditions, including the initial state of the atom and the electromagnetic field, i.e.,
\begin{equation}\label{C I}
   |\psi(0)\rangle = |\psi(0)\rangle_{\text{atom}} \otimes |\psi(0)\rangle_{\text{field}}.
\end{equation}

\subsection{Fourth-order Runge-Kutta Method}
RK4 was originally developed in 1895 by the physicist and mathematician Carl David Runge for the numerical solution of differential equations, and was extended in 1901 by M. Wilhelm Kutta to systems of differential equations. This method is an iterative procedure that approximates the solution to a differential equation through a series of discrete steps. Unlike simpler methods, such as Euler’s method, RK4 provides a fourth-order approximation, meaning the error per step is proportional to $h^5$, where $h$ is the step size in time~\cite{Butcher_book,Press_book}.

\begin{figure}[H]
\centering
\begin{overpic}[width=0.5\linewidth]{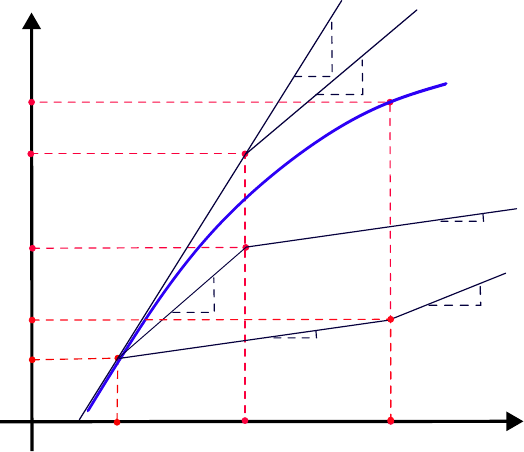}
        \put(0,0){ \Large $0$}
        \put(18,0){ \large $x_i$}
        \put(38.7,0){ \large $x_i+\frac{h}{2}$}
        \put(66.5,0){ \large $x_i + h$}
        \put(95,0){ \Large $x$}
        \put(-5,17){ \large $y_i$}
        \put(-12,24){ \large $y_i+hk_3$}
        \put(-11,38){\large $y_i+\frac{hk_2}{2}$}
        \put(-11,56){\large $y_i+\frac{hk_1}{2}$}
        \put(-7,66.5){ \large $y_{i+1}$}
        \put(1.5,85){ \Large $f(x)$}
        \put(63,76){ \large $k_1$}
        \put(69,71){ \large $k_2$}
        \put(40.5,29){ \large $k_2$}
        \put(92,42){ \large $k_3$}
        \put(60,19.5){ \large $k_3$}
        \put(92,28.5){ \large $k_3$}
\end{overpic}
\caption{Geometric interpretation of the fourth-order Runge-Kutta method.}
\label{fig RK4}
\end{figure}
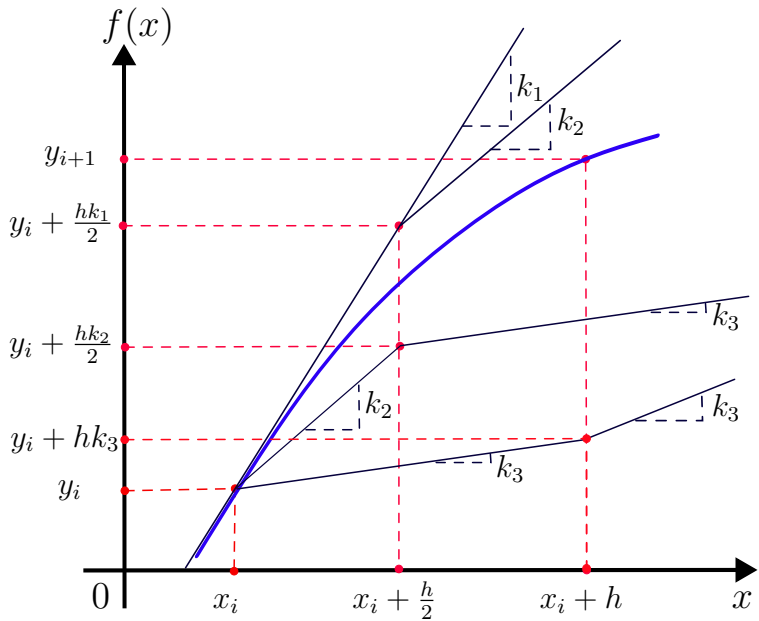

In general, the numerical solution of the initial value problem $y' = f(x,y)$, with $y(x_0) = y_0$ and step size $h$, is obtained through the points $(x_{i+1}, y_{i+1})$, computed using the following recurrence formulas:
\begin{align*}
x_{i+1} &= x_i + h, \\
k_1 &= f(x_i, y_i), \\
k_2 &= f\left(x_i + \frac{h}{2}, y_i + h \frac{k_1}{2} \right), \\
k_3 &= f\left(x_i + \frac{h}{2}, y_i + h \frac{k_2}{2}\right), \\
k_4 &= f(x_i + h, y_i + h k_3), \\
y_{i+1} &= y_i + \frac{h}{6}(k_1 + 2 k_2 + 2 k_3 + k_4),
\end{align*}
where $i=0,1,2,\ldots$. The sum $\left( k_1 + 2k_2 + 2k_3 + k_4 \right)/6$ can be interpreted as an average slope. Note that $k_1$ is the slope at the beginning of the interval, $k_2$ is the slope at the midpoint approaching $x_i + \frac{h}{2}$, $k_3$ is a second estimate of the slope at the same midpoint, and $k_4$ is the slope at $x_{i} + h$. The geometric interpretation of this RK4 scheme can be seen in figure \ref{fig RK4}.

In the case of the Schrödinger equation~\eqref{Ecuacion de Schrodinger}, it is clear that $x_i = t_i$, i.e., time, and $y_i = |\psi_i\rangle$, which corresponds to the quantum state of the system.

The main advantage of RK4 is that it is easy to program, making it an attractive option for the numerical solution of differential equations. However, a significant disadvantage is the need to evaluate the function $f(x,y)$ at multiple different points in $x$ and $y$ during each step. This can increase computational time, especially in problems where evaluating $f(x,y)$ is costly or for systems with high dimensionality. Despite this drawback, RK4 remains widely used because of its balance between accuracy and simplicity of implementation.

\subsection{Implementation in Python}

Next, the code implementing RK4 to solve the Schrödinger equation~\eqref{Ecuacion de Schrodinger}, associated with the Hamiltonian describing the system~\eqref{Hamiltonian JCM}, is presented. It is worth noting that the Python interface is set to English, which restricts the use of accented characters and special symbols. For this reason, any word appearing without accents within the code is not a spelling mistake but a limitation imposed by the programming environment.

\begin{enumerate}
\item \textbf{Importing libraries} \\
In this first step, the necessary libraries for performing numerical operations, matrix manipulation, and graph generation are imported. These libraries are fundamental for defining quantum operators, solving the equations, and visualizing the simulation results.
\begin{lstlisting}[style=python, caption={}]
import numpy as np
from scipy.special import factorial
import matplotlib.pyplot as plt
\end{lstlisting}
\noindent
\begin{itemize}
    \item \texttt{numpy}: A widely used library in Python for creating and manipulating arrays and matrices. It is essential for defining the quantum operators of the system.
    \item \texttt{scipy.special.factorial}: This function from \texttt{scipy} calculates the factorial of a number, which is key for expressing Fock states (discrete states of the quantum field) in the simulation.
    \item \texttt{matplotlib.pyplot}: A library for generating plots. It will allow us to visualize the time evolution of quantum observables, such as the atomic inversion or the average number of photons in the field.
\end{itemize}

\item \textbf{Defining system constants} \\
In this step, the fundamental parameters of the model are specified, representing the physical properties of the quantum system.
\begin{lstlisting}[style=python, caption={}]
N        = 50   
omega_c  = 0.4   
omega_eg = 0.9  
g        = 1.0  
hbar     = 1.0  
t_f      = 30   
dt       = 0.05 
\end{lstlisting}
\noindent
\begin{itemize}
    \item \texttt{N}: Represents the maximum number of Fock states considered in the system.
    \item \texttt{omega\_c}: The frequency of the electromagnetic field.
    \item \texttt{omega\_eg}: The transition frequency between the energy levels of the atom.
    \item \texttt{g}: The coupling constant, which determines the strength of the interaction between the atom and the field.
    \item \texttt{hbar}: The reduced Planck constant ($\hbar$), here set to 1 to simplify units.
    \item \texttt{t\_f}: The final time of the simulation, i.e., until when the system's evolution will be calculated.
    \item \texttt{dt}: The time step used in the numerical evolution (previously denoted as $h$). The smaller this value, the higher the simulation's accuracy, though at a higher computational cost.
\end{itemize}

\item \textbf{Definition of the operators and the Hamiltonian modeling the system} \\
In this step, the creation and annihilation operators are defined, as well as the Pauli operators acting on the system's Hilbert space (to see the matrix form of these operators, it is necessary to consult some of the following references~\cite{Gerry_Book, Klimov_2009, Garrison_2008, Agarwal_2013, Fox_2006, Hernandez_book}). From these operators, the Hamiltonian of the Jaynes-Cummings model is constructed, which describes the interaction between the electromagnetic field and the atom.
\begin{lstlisting}[style=python, caption={}]
def destroy(N):
    return np.diag(np.sqrt(np.arange(1, N)), 1) 

a = destroy(N)
ad = a.T.conj()
sigma_z = np.array([[1, 0], [0, -1]])
sigma_p = np.array([[0, 1], [0, 0]])
sigma_m = np.array([[0, 0], [1, 0]])

def tensor(A, B):
    return np.kron(A, B)         
 
def H_JCM(omega_c, omega_eg, g): 
    return (omega_c * tensor(np.dot(ad, a), np.eye(2)) + 
            (omega_eg / 2) * tensor(np.eye(N), sigma_z) + 
            g * (tensor(a, sigma_p) + tensor(ad, sigma_m)))    
\end{lstlisting}
\noindent
\begin{itemize}
    \item \texttt{destroy(N)}: This function creates the annihilation operator of dimension $N$ in matrix form, using \texttt{np.diag} to generate a diagonal matrix containing square roots of integers, shifted by 1 above the main diagonal.
    \item \texttt{a}: Represents the annihilation operator $\hat{a}$, defined from the function \texttt{destroy}.
    \item \texttt{ad}: Is the creation operator $\hat{a}^\dagger$, obtained as the conjugate transpose of the annihilation operator $\hat{a}$.
    \item \texttt{sigma\_z}: Defines the Pauli operator $\hat{\sigma}_z$ in matrix form, representing whether the atom is in the excited or ground state.
    \item \texttt{sigma\_p} and \texttt{sigma\_m}: The Pauli operators are $\hat{\sigma}_+$ and $\hat{\sigma}_-$, which represent the upward and downward transitions between the atom's energy levels, respectively.
    \item \texttt{tensor(A, B)}: This function calculates the tensor product of matrices $A$ and $B$ using \texttt{np.kron}, allowing the combination of the atom and field Hilbert spaces.
    \item \texttt{H\_JCM(omega\_c, omega\_eg, g)}: This function defines the Hamiltonian of the Jaynes-Cummings model~\eqref{Hamiltonian JCM}, which includes terms for the electromagnetic field, the atomic transitions, and their interaction. Tensor product operations and matrix multiplication are used to construct the full Hamiltonian.
\end{itemize}

\item \textbf{Definition of the Schrödinger equation} \\
In this step, the time-dependent Schrödinger equation is formulated, which describes the evolution of the system’s quantum state under the influence of the previously defined Hamiltonian.
\begin{lstlisting}[style=python, caption={}]
def Schrodinger(t, psi, H):
    return -1j / hbar * np.dot(H, psi)
\end{lstlisting}
\noindent
\begin{itemize}
    \item \texttt{Schrodinger(t, psi, H)}: This function implements the Schrödinger equation, which relates the temporal evolution of the quantum state $\ket{\psi}$ to the system's Hamiltonian.
    \item \texttt{t}: Represents the time at which the quantum state evolution is evaluated.
    \item \texttt{psi}: Is the state vector of the system at a given instant, which is modified as the simulation progresses.
    \item \texttt{H}: Is the Hamiltonian of the system, which acts on the quantum state $\ket{\psi}$ to determine how it evolves over time.
    \item The Schrödinger equation is presented in matrix form, where the system's Hamiltonian acts on the state vector $\ket{\psi}$. The result is multiplied by $-i/\hbar$ (where \texttt{1j} in Python represents the imaginary unit $i$), and $\hbar$ is the reduced Planck constant. This process allows calculating the rate of change of the quantum state with respect to time.
\end{itemize}

\item \textbf{Implementation of RK4} \\
In this step, RK4 is used to numerically solve the Schrödinger equation. This method allows calculating the system's quantum state $\ket{\psi(t)}$ over time by dividing the total time into small discrete steps and approximating the solution step-by-step.

\begin{lstlisting}[style=python, caption={}]
def RK4(psi_0, H, t_list):  
    psi_t = np.zeros((len(t_list), len(psi_0)), dtype=complex)  
    psi_t[0] = psi_0

    for i, t in enumerate(t_list[:-1]):  
        k1 = Schrodinger(t, psi_t[i], H)  
        k2 = Schrodinger(t + dt / 2, psi_t[i] + k1 * dt / 2, H)
        k3 = Schrodinger(t + dt / 2, psi_t[i] + k2 * dt / 2, H)  
        k4 = Schrodinger(t + dt, psi_t[i] + k3 * dt, H) 
        
        psi_t[i + 1] = psi_t[i] + (dt / 6) * (k1 + 2 * k2 + 2 * k3 + k4)
    
    return psi_t
\end{lstlisting}

Next, each step of the code is explained:

\begin{itemize}
    \item \texttt{psi\_t}: This creates an array filled with zeros to store the quantum states at each time point. The array size corresponds to the number of time steps specified, and each element has the same dimension as the initial state \texttt{psi\_0}. This array is of complex data type because quantum states involve complex numbers.
    \item \texttt{psi\_t[0]}: This sets the initial quantum state \texttt{psi\_0} at time zero.
    \item \texttt{for i, t in enumerate(t\_list[:-1])}: 
    This loop iterates over all time points in \texttt{t\_list} except the last one. In each iteration, it computes the quantum state at the next time step.
    \item \texttt{k1}: 
    The initial estimate at the start of the step. It evaluates the Schrödinger equation at current time \texttt{t} with the current state \texttt{psi\_t[i]}.
    \item \texttt{k2}: 
    The estimate at the midpoint, calculated using \texttt{k1} to update the state, advancing by half a step. It evaluates the Schrödinger equation at time \texttt{t + dt/2}.
    \item \texttt{k3}: 
    Similar to \texttt{k2}, but uses \texttt{k2} for the midpoint estimate; again evaluated at time \texttt{t + dt/2}.
    \item \texttt{k4}: 
    The estimate at the full step, using the \texttt{k3} value to update the state, evaluated at time \texttt{t + dt}.
    \item \texttt{psi\_t[i + 1]}: 
    The next quantum state is obtained by combining these four estimates according to the RK4 formula, which provides a weighted average to improve accuracy.
\end{itemize}

\item \textbf{Definition of the initial state of the atom-field system} \\
In this section, the initial quantum state of the system, composed of the electromagnetic field and the atom, is defined. The field state can be represented by various types of quantum states, such as Fock states, coherent states, or squeezed states, among others. For the atom, initial conditions can include the ground state, the excited state, or a superposition of both.

For this implementation, a coherent state $|\alpha\rangle$ for the field and the excited state $|e\rangle$ for the atom have been chosen as initial conditions. However, this formalism is sufficiently general to be applied to other types of states for both the field and the atom.

\begin{lstlisting}[style=python, caption={}]
alpha = 3.0 
psi_field = np.exp(-np.abs(alpha)**2 / 2) * np.array([alpha**n / np.sqrt(factorial(n)) for n in range(N)])  
psi_atom = np.array([1, 0])
psi_0 = np.kron(psi_field, psi_atom) 
\end{lstlisting}
\noindent
\begin{itemize}
    \item \texttt{alpha}: 
    Defines the parameter \texttt{alpha}, which characterizes the coherent state of the field. This parameter is related to the mean amplitude of the electromagnetic field.
    \item \texttt{psi\_field}: 
    This line mathematically defines the coherent state of the field.
    \item \texttt{psi\_atom}: 
    Here, the initial state of the atom is defined. In this case, \texttt{np.array([1, 0])} indicates that the atom is in the excited state $\ket{e}$.
    \item \texttt{psi\_0}: 
    The initial state of the entire atom-field system is obtained by the tensor product (\texttt{np.kron}) of the coherent field state and the atom state. The tensor product combines the two subsystems into a single quantum state that describes the entire system.
\end{itemize}

\item \textbf{Calculation of the system’s quantum state time evolution and the extraction of any observable} \\
In this step, RK4 is used to solve the time-dependent Schrödinger equation. Using the previously defined system Hamiltonian, the time evolution of the atom-field quantum state is computed by evaluating the Schrödinger equation solution at a list of discrete times.

Once the state vector $\ket{\psi(t)}$ is obtained, it is possible to calculate any quantum observable of the system. In this case, the atomic inversion $W(t)$ has been chosen, which corresponds to the expected value of the operator $\hat{\sigma}_z$. This observable provides crucial information about the population difference between the atom's ground and excited states, allowing characterization of the atom's dynamics in interaction with the electromagnetic field.

\begin{lstlisting}[style=python, caption={}]
t_list = np.arange(0, t_f, dt)
H = H_JCM(omega_c, omega_eg, g)
psi_t = RK4(psi_0, H, t_list)
W_t = np.real(np.array([np.vdot(psi, np.dot(tensor(np.eye(N), sigma_z), psi)) for psi in psi_t]))
\end{lstlisting}

\begin{itemize}
    \item \texttt{t\_list}: A time range is defined from \texttt{0} to \texttt{t\_f}, with a time step of \texttt{dt}. These values determine the points at which the quantum state will be evaluated during the time evolution.
    \item \texttt{H}: Based on the atom and field frequencies (\texttt{omega\_c}, \texttt{omega\_eg}) and the coupling constant \texttt{g}, the Hamiltonian of the Jaynes-Cummings model is constructed, which governs the system's dynamics.
    \item \texttt{psi\_t}: Starting from the initial state \texttt{psi\_0} and the Hamiltonian \texttt{H}, the Schrödinger equation is solved over time using RK4, yielding the system's evolution at the specified times in \texttt{t\_list}.
    \item \texttt{W\_t}: Calculates the expected value of the atomic inversion at each time instant.
\end{itemize}
\end{enumerate}

\subsection{Method validation}

Once the state vector $\ket{\psi(t)}$ is obtained, the atomic inversion $W(t)$ has been calculated as an example of an observable. To validate that the code implemented in the methodology correctly analyzes the system’s time evolution, we have plotted the atomic inversion obtained numerically via RK4, comparing it with the exact solution for the atomic inversion in the Jaynes-Cummings model under the same initial conditions. The exact solution is given by~\cite{Hernandez_book}:
\begin{equation}
    W(t) =  \sum_{n=0}^\infty \frac{P_n}{\Omega_{n+1}^2} \left[  \left( \frac{\omega_c -\omega_{eg}}{2}\right)^2 + g^2 (n+1) \cos \left( 2 \Omega_{n+1} t \right) \right],
\end{equation}
where $\Omega_{n+1}^2 = \sqrt{\left( \frac{\omega_c - \omega_{eg}}{2}\right)^2 + g^2 (n+1)}$ is the Rabi frequency, and $P_n = e^{-\bar{n}} \frac{\bar{n}^n}{n!}$ is the probability distribution associated with a coherent state. The exact expression in Python can be written as follows:

\begin{lstlisting}[style=python, caption={}]
def DPE_EC(n, alpha):        
    return np.exp(-abs(alpha)**2) * abs(alpha)**(2*n) / factorial(n)
def Omega_n1(n, omega_c, omega_eg, g):    
    return np.sqrt(((omega_c - omega_eg) / 2)**2 + g**2 * (n+1))
def Parentesis_n1(t, n, omega_c, omega_eg, g):   
    return ((omega_c - omega_eg) / 2)**2 + g**2 * (n+1) * np.cos(2 * Omega_n1(n, omega_c, omega_eg, g) * t)
def W_JCM(t, alpha, omega_c, omega_eg, g, N):
    return sum(DPE_EC(n, alpha) * 
    Parentesis_n1(t, n, omega_c, omega_eg, g) / (Omega_n1(n, omega_c, omega_eg, g)**2) for n in range(0, N))
\end{lstlisting}
\begin{itemize}
    \item \texttt{DPE\_EC(n, alpha)}: This function calculates the probability distribution $P_n$.
    \item \texttt{Omega\_n1(n, omega\_c, omega\_eg, g)}: Computes the Rabi frequency $\Omega_{n+1}$, which depends on the field frequency \texttt{omega\_c}, the atom frequency \texttt{omega\_eg}, and the coupling constant \texttt{g}.
    \item \texttt{Parentesis\_n1(t, n, omega\_c, omega\_eg, g)}: Evaluates the term inside the brackets in the formula for $W(t)$.
    \item \texttt{W\_JCM(t, alpha, omega\_c, omega\_eg, g, N)}: Finally, this function performs the summation to calculate the atomic inversion $W(t)$.
\end{itemize}

Next, we illustrate how to plot both the atomic inversion computed numerically using the Runge-Kutta method (solid line) and the exact solution (stars):

\begin{lstlisting}[style=python, caption={}]
W_numeric = np.array([np.real(np.vdot(psi, np.dot(tensor(np.eye(N), sigma_z), psi))) for psi in psi_t])
step = 10  
t_exact_list = t_list[::step]
W_exact = np.array([W_JCM(t, alpha, omega_c, omega_eg, g) for t in t_exact_list])


plt.plot(t_list, W_numeric, label="RK4", color="blue", linestyle="-")
plt.plot(t_exact_list, W_exact, label="Exact", color="red", marker="*", linestyle="")

plt.xlabel(r'Time $t$', fontsize=14)
plt.ylabel(r'$W(t)$', fontsize=14)
plt.title('RK4 vs Exact', fontsize=16)
plt.grid(True)
plt.legend()

plt.ylim(-1, 1)
plt.yticks(np.arange(-1, 1.5, 0.5))
plt.xticks(fontsize=12)
plt.yticks(fontsize=12)

plt.savefig('atomic_inversion.pdf', format='pdf')
plt.show()
\end{lstlisting}

In the previous code:
\begin{itemize}
    \item \texttt{step}: A step size of 10 is defined to space out the stars in the exact solution plot more widely. This way, fewer points are selected from the list of times \texttt{t\_list}, making the stars appear more separated visually.
    \item \texttt{t\_exact\_list}: A new list of times \texttt{t\_exact\_list} is generated, containing only every tenth element from the original \texttt{t\_list}, to spread out the markers in the exact solution plot.
    \item \texttt{W\_exact}: Calculates the exact atomic inversion using the previously defined function \texttt{W\_JCM}, for each time in the list \texttt{t\_exact\_list}.
    \item \texttt{plt.plot()}: Both results are plotted. The numerical calculation (\texttt{W\_numeric}) is shown as a continuous blue line, while the exact solution (\texttt{W\_exact}) is represented with red stars to mark the discrete values.
    \item \texttt{plt.xlabel}, \texttt{plt.ylabel}, \texttt{plt.title}: Set the labels and title of the plot, with font size adjustments.
    \item \texttt{plt.grid(True)}: Activates a grid on the plot to make it easier to visualize the points and lines.
    \item \texttt{plt.legend()}: Displays the legend to distinguish between the numerical results (RK4) and the exact solution.
    \item \texttt{plt.ylim(-1, 1)}: Sets the y-axis limits between -1 and 1 so that the values of the atomic inversion fit within this range.
    \item \texttt{plt.yticks(np.arange(-1, 1.5, 0.5))}: Defines the y-axis tick labels with intervals of 0.5 from -1 to 1, for improved readability.
    \item \texttt{plt.xticks}, \texttt{plt.yticks}: Adjust the font sizes of the x and y axis labels for better visibility.
    \item \texttt{plt.savefig()}: Saves the plot in any format, in this case as a PDF file, allowing for high-quality inclusion in documents or presentations.
\end{itemize}

The resulting plot looks like this:
\begin{figure}[H] 
    \centering 
    \includegraphics[width=0.6\linewidth]{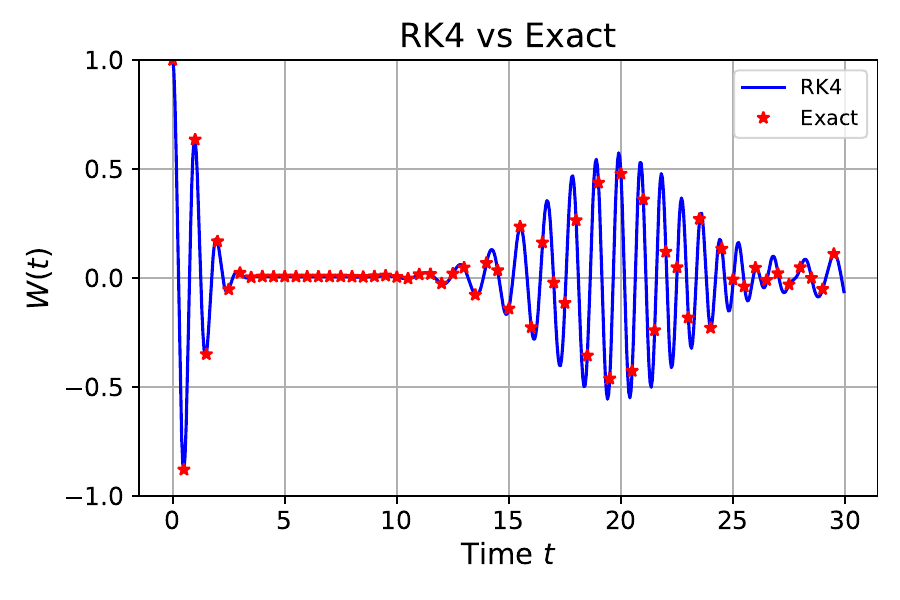}
    \caption{Graph of atomic inversion in the Jaynes-Cummings model. The solid blue line represents the numerical solution obtained using the Runge-Kutta method, while the red stars show the exact solution.}
    \label{Fig2}
\end{figure}

\section{Results and Discussion}\label{Resultados}

The results obtained through the numerical implementation of RK4 to solve the time-dependent Schrödinger equation show a remarkable consistency with the exact solution of the Jaynes-Cummings model. The comparative plot of the atomic inversion, contrasting both solutions, validates the accuracy of the numerical approach (Figure \ref{Fig2}), faithfully reproducing the characteristic oscillatory behavior of the atom-field system under the initial conditions imposed, as expected in the theoretical models.

The fact that the numerical results coincide with the exact solution confirms that the employed methodology is robust and accurate for this type of quantum problems. Moreover, the use of RK4 offers significant advantages, such as its capacity to handle more complex Hamiltonians, where exact solutions are not available or are difficult to obtain.

This approach is not limited to the Jaynes-Cummings model. Since the method is completely general, only the Hamiltonian modeling the system needs to be modified to analyze other radiation-matter interaction scenarios. For example, in situations involving multiple field modes or more complex interactions~\cite{Gerry_Book, Hernandez_book}, the procedure remains the same: simply define the relevant operators, and then formulate the new Hamiltonian that models the system under study, as detailed in step 3 of the methodology presented in Section \ref{Methodology}.

\section{Conclusions}\label{Conclusions}
In this article, we have presented a numerical methodology based on the fourth-order Runge-Kutta method to solve the time-dependent Schrödinger equation in quantum systems of radiation-matter interaction. Using the Jaynes-Cummings model as an example, it has been demonstrated that the numerically obtained results are consistent with the known exact solutions, thus confirming the effectiveness of the method.

The flexibility of the proposed approach allows its application to a wide variety of quantum systems. It is only necessary to adapt the system's Hamiltonian and, if needed, define new operators to analyze the relevant observables. This makes the method a powerful and adaptable tool for studying complex quantum systems where exact solutions are not available or are very difficult to obtain.

Finally, using programming languages like Python is ideal, since it is open-source, easily accessible, and highly efficient. This makes it an excellent alternative for all types of research.

\section*{Acknowledgements}
L. Hernández-Sánchez thanks the Instituto Nacional de Astrofísica, Óptica y Electrónica (INAOE) for the collaboration scholarship granted and the Consejo Nacional de Humanidades, Ciencias y Tecnologías (CONAHCYT) for the SNI Level III assistantship (CVU No. 736710).

\bibliographystyle{unsrt} 
%


\end{document}